\documentclass[]{spie}  

\usepackage{amsmath,amsfonts,amssymb}
\usepackage{graphicx}
\usepackage[colorlinks=true, allcolors=blue]{hyperref}

\title{InfraRed On-Detector Guide-Windows in the era of Extremely Large Telescopes}

\author[a]{Edward L. Chapin$^*$}
\author[a]{Jennifer Dunn}
\author[a]{Tim Hardy}
\author[a]{Owen Hubner}
\author[a]{Jordan Lothrup}

\affil[a]{National Research Council Herzberg, 5071 W Saanich Rd, Victoria, V9E 2E7, Canada}

\authorinfo{$^*$ed.chapin@nrc-cnrc.gc.ca, phone +1 250 363 6971}

\begin{document}
\maketitle

\begin{abstract}
Future Extremely Large Telescopes (ELTs) will require advances in Adaptive Optics (AO) systems to fully realize their potential. In addition to separate, dedicated wavefront sensors, it is recognized that wavefront sensing within the science focal plane itself will also be needed for many new instruments. One approach is to use On-Detector Guide Windows (ODGWs), whereby a small sub-window of a science detector is read-out continuously ($\sim$10s-100s of Hz) in parallel with slower reads of the full chip ($>$10\,s). Guide star centroids from these windows can be used to correct for vibrations and flexure. Another potential use for these windows is to perform localized resets at high cadence to prevent saturation and to minimize persistence from bright sources. We have prototyped an ODGW system using a 5-$\mu$m cutoff Teledyne HAWAII-2RG infrared detector, and the new Astronomical Research Cameras Gen-4 controller. We describe our implementation of an ODGW mode, and science image artifacts that were observed.
\end{abstract}


%
%

\section{INTRODUCTION}
\label{sec:intro}

In this paper we report on prototyping results for a hybrid full-frame and ``On-Detector Guide Window'' (ODGW) observing mode for Teledyne HAWAII-xRG infrared detectors using a new Generation 4 (Gen-4) controller from Astronomical Research Cameras (ARC)\footnote{http://www.astro-cam.com/Gen4Hardware.php}. Our laboratory system is based around a 5-$\mu$m cutoff HAWAII-2RG, and captures key aspects of the ODGW design proposed for the InfraRed Imaging Spectrograph (IRIS) first-light camera currently being developed for the Thirty Meter Telescope (TMT)\cite{larkin2020,dunn2016}. A similar concept is also being explored for the Gemini InfraRed Multi-Object Spectrometer (GIRMOS)\cite{chapman2018} imager subsystem. The general idea is to provide frequent visits to one or more sub-windows (e.g., 10s to 100s of Hz) in parallel with slower reads of the entire chip used for science. These smaller windows would normally target guide stars to provide fast tip/tilt feedback to an adaptive optics (AO) system. The two key benefits of ODGWs over external low-order wavefront sensors are: (i) measurements are provided directly from the science focal plane (i.e., they can sense subtle time-varying flexure between the science camera and other sensors); and (ii) they can extend the sky-coverage of the instrument (since guide stars may be selected from within the science detector field-of-regard). Other potential uses for independently-addressable sub-windows are to rapidly read and reset regions around particularly bright sources, with the goal of mitigating localized persistence effects, and to provide high-cadence measurements of time-varying astronomical sources.

The hybrid full-frame plus fast sub-window concept is not new, and our particular implementation is very similar to the ``Guide Mode'' described in HAWAII-1RG technical documentation. The full detector is traversed line-by-line. At the end of each row it is temporarily switched into window readout mode, and it clocks through all of the window pixels. It then reverts to the normal readout mode and continues with the next row of the full detector, and so on. In this way a single window can be visited up to NROWS (= the total number of rows in the detector) times faster than the full frame, although each row will take slightly longer to accommodate the extra pixels in the window.

Despite the potential utility of this HAWAII-xRG mode of operation, it does not seem to be used extensively, with a literature search yielding only a handful of relevant examples. An on-chip guiding system has been used since the early 2000s for the Wide-field InfraRed Camera (WIRCAM) on the Canada France Hawaii Telescope (CFHT), with a mosaic of four HAWAII-2RG near-IR detectors \cite{loic2005,baril2006,vermeulen2006}. However, WIRCAM uses a completely separate controller from the science data path to acquire the guide window pixels, and the system switches between normal and window mode for complete exposures (rather than interleaving them). The Carnegie Astrometric Planet Search Cameras (CAPScam) uses the window feature of HAWAII-2RG HyViSI optical detectors in a manner more similar to what is described here to rapidly read and reset bright stars \cite{boss2009}. An ODGW mode was implemented for the Gemini South Adaptive-Optics Imager (GSAOI) which is fed by the Gemini GeMS multi-conjugate adaptive optics system \cite{young2012}. Other studies that provide useful context include: the use of windows on a HAWAII-1RG detector for a radial velocity experiment \cite{bezawada2006}, and fast window readouts from a HAWAII-2RG for a dedicated near-infrared tip/tilt sensor \cite{smith2012}. These past efforts mention various idiosyncrasies, and so the ODGW concept is still seen as a risk to both IRIS and GIRMOS.

While the IRIS imager will ultimately be built by collaborators at Caltech, Herzberg Astronomy and Astrophysics (HAA) was in a position to perform some early prototyping of the technique due to the fact that an ARC Gen-4 controller, the current baseline for IRIS, was recently purchased for another project, and we were already in possession of several HAWAII-xRG detectors. HAA is also building the IRIS On-Instrument Wavefront Sensor (OIWFS), and the Narrow Field InfraRed Adaptive Optics System (NFIRAOS) which feeds IRIS, so there is a strong interest in prototyping the real-time integration of the ODGW with the rest of the AO system. Finally, HAA will also be building the GIRMOS imager subsystem\cite{atwood2022}. We note that Caltech has also been conducting research into ODGWs, building on past experiments with fast windowed readouts\cite{smith2012}. As members of the IRIS consortium we have been privy to their ongoing efforts, and we specifically acknowledge the support of Roger Smith and Tim Greffe who provided results from their own tests, and offered helpful suggestions for our system.

%
%

\section{Test system description}
\label{sec:config}

\begin{figure}[hbt]
    \begin{center}
        \includegraphics[width=0.9\linewidth]{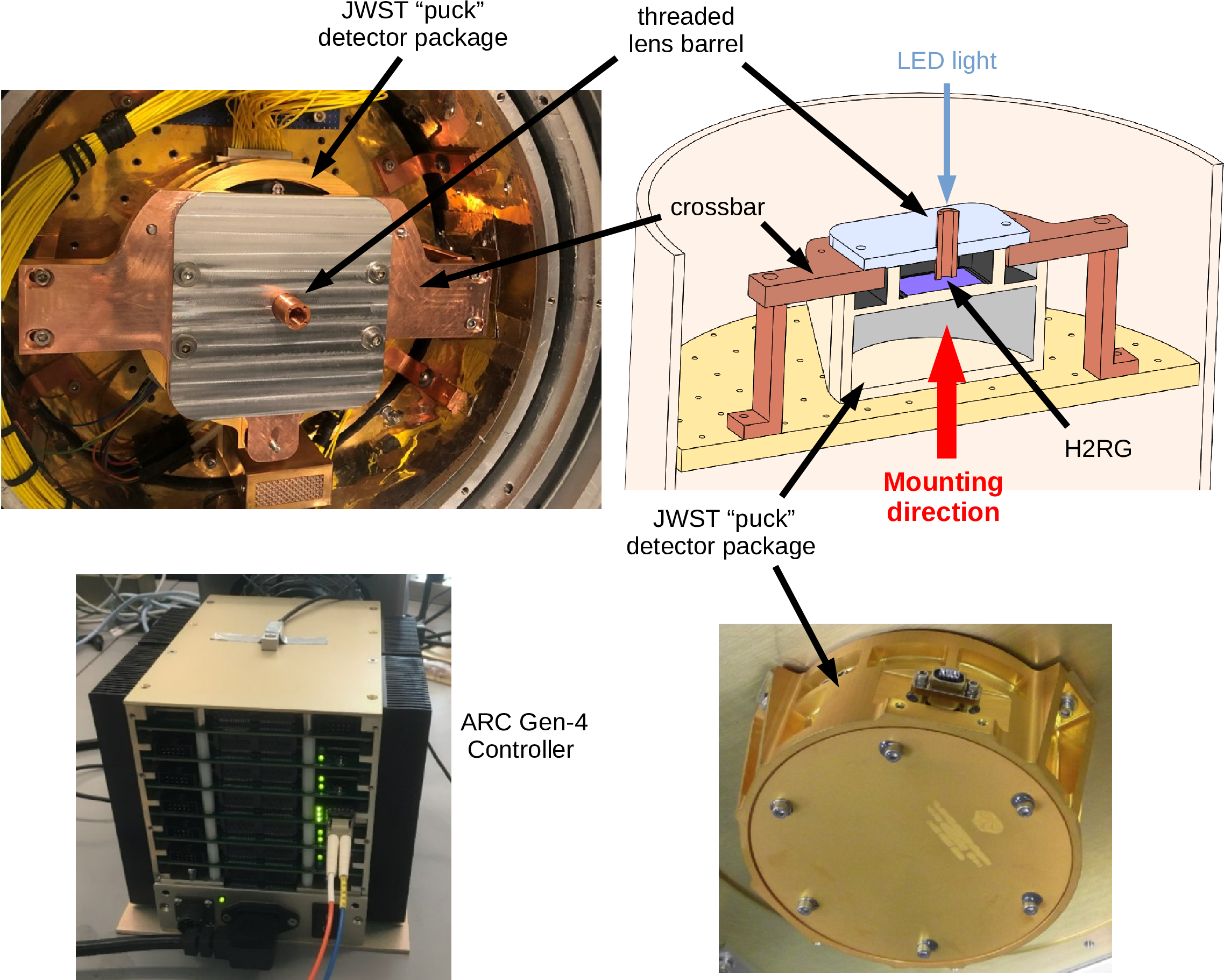}
    \end{center}  
    \caption{\label{fig:cryostat}
    Test system, including the cryostat with detector and light source mounts, and the ARC Gen-4 controller. The JWST package was designed to mount to a flange above the detector. The photo in the lower-right shows the bottom of the package. To accommodate this orientation we fabricated a crossbar (copper) with legs from which the package is suspended. The detector is near the top of the package, enshrouded in a shallow baffle. An aluminum cover plate further prevents stray light. An optional threaded barrel with a lens at the bottom passes through the plate, and is used to manually focus light produced by an infrared LED that is mounted at the top of the barrel (within the cryostat). Otherwise the barrel can be replaced with a simple pinhole to enable broad illumination of the detector by the LED.
    }
\end{figure}

Our prototype system uses a 5-$\mu$m cutoff HAWAII-2RG 2k $\times$ 2k sensor, with an 18\,$\mu$m pixel pitch. Four rows of reference pixels at the top and bottom, and four columns of reference pixels on the sides can be used to track systematic effects. This particular detector was originally purchased by the Canadian Space Agency (CSA) for the James Webb Space Telescope (JWST) Fine Guidance Sensor / Near InfraRed Imager and Slitless Spectrograph. However, when it was discovered that it suffered from a serious indium diffusion problem\cite{rauscher2012} it was offered to HAA for engineering experiments.

%
%
%

The cryostat selected for our experiment can be filled with liquid cryogens, and has two radiation shields, both of them connected to a $^4$He closed-cycle cryo-cooler. The radiation shields were heat-strapped to the cold plate for a previous project. It can thus be pre-cooled using liquid nitrogen (LN2), and then operated continuously with the cold head down to temperatures reaching $\sim$40\,K. For most of our experiments the system was not controlled, and detector temperatures typically varied between $\sim$45--50\,K peak-to-peak over a 24-hr period. Both the inner and outer radiation shields have holes to allow light to enter through an external window. For these experiments the hole in the outer shield is covered with metal tape to minimize thermal stray light from the room. A synthetic light source occupies the hole in the inner radiation shield, as described below.

The detector itself is mounted in its JWST flight package. While this package offers some desirable stray-light baffling, it also presented a challenge due to the fact that it mounts to a flange on the outward facing side of the detector. For our initial tests we therefore attached the package to the cold plate face-down. Once we determined that it was working, we fabricated a new mount as shown in Figure~\ref{fig:cryostat}, that allowed us to point the detector outward, and to illuminate it with an LED.

The Gen-4 controller consists entirely of warm components. While its general architecture is inspired by the earlier Gen-3 system, it was re-designed from the ground-up using modern components. The controller itself consists of a stack of boards in a custom enclosure including: a timing board that implements the main control and communication signals; a clock driver and DC bias board; one to four video processor boards, each with 16 DC-coupled channels (providing a maximum of 64 channels) that feed 16-bit analog-to-digital converters (ADCs) with a peak signal amplitude of 0.68\,V at the input to the board, and with individually programmable offsets (though fixed gain); and finally a built-in power supply. Board logic is implemented with FPGAs, and firmware updates from ARC can be easily deployed in the field. In our configuration the digitized values recorded by the system are inverted; i.e., a smaller number of analog-to-digital units (ADUs) corresponds to a larger accumulated charge in a pixel.

When the controller was first delivered to HAA in 2020 it was connected via a fiber from the timing board to a standard 10-GbE network interface controller (NIC). However, we encountered difficulties with this interface (such as dropped packets). In 2021 ARC decided to abandon this approach, and now exclusively supports a separate interface using a proprietary PCIe card. Much like the earlier Gen-3 controller, the Linux network stack is bypassed, and instead a dedicated kernel module is used for communication.

The controller's activities are orchestrated by an ARM microcontroller that is programmed in C. All of the source code is provided by ARC, including a base library and example code to set the biases, continuously execute idle waveforms while the detector is not exposing (e.g., to continuously reset the sensor), and perform Fowler-sampled exposures when requested. A C++ application programming interface (API), and a Linux kernel module (C) are also provided, including the full source code. The experiments described here began with a snapshot of the ARC software from November 2021 (API version 4.5.9). All of the server-side and microcontroller code development is performed on the host PC. An Eclipse-based IDE (MCUXpresso) is used for the microcontroller development. The resulting compiled binaries are then uploaded to the controller on startup using a C++ API call.  

The HAWAII-2RG is operated with a single video output channel (primarily due to limitations of existing cryogenic wiring from a previous experiment), and uses the slow (100\,kHz) output mode. The biases were set to $V_\mathrm{RESET}=0.3$\,V, $V_\mathrm{DSUB}=0.55\,V$, $V_\mathrm{DD}=3.3$\,V, and $V_\mathrm{BIASGATE}=2.23$\,V. The overall gain of the system is estimated to be 3.3\,e-/ADU, based on the output sensitivity (3.17\,$\mu$V/e-, from device documentation), the controller gain of 12 prior to the ADC, and the ADC conversion factor of 8000\,ADU/V.

\begin{figure}[hbt]
    \begin{center}
        \includegraphics[width=0.9\linewidth]{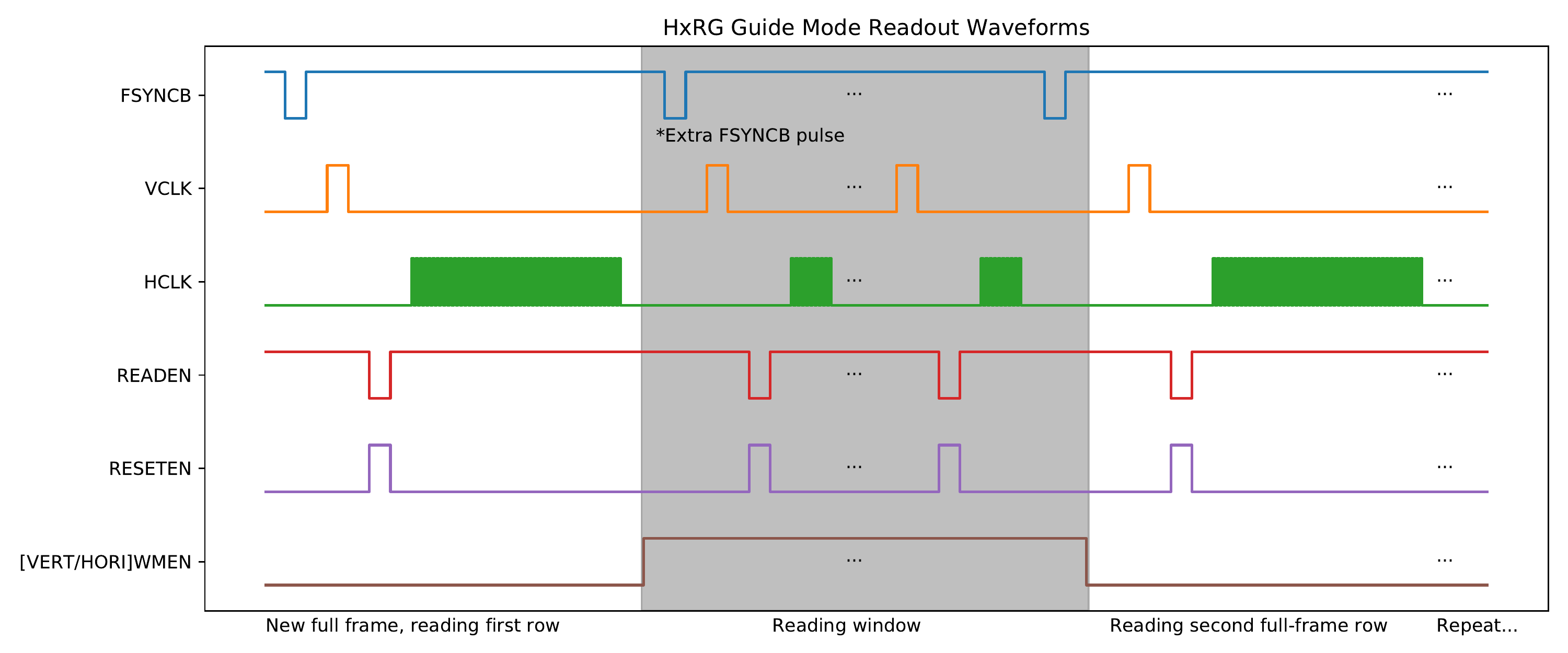}
    \end{center}  
    \caption{\label{fig:waveform}
    Illustration of the timing signals used to implement the hybrid full-frame and ODGW observing mode. Note the extra FSYNCB pulse which is added to the beginning of the window reading sequence. In addition, serial commands are sent to switch into and out of window mode (not shown). In this example reset pulses are used both for the full-frame and in the window. The full-frame and window reset behaviors are independently configured.
    }
\end{figure}

We implemented the ODGW exposure mode in software as a fork of the basic Fowler-sampled full-frame mode provided by ARC. Instead of performing a reset, executing $N_1$ reads at the beginning, waiting for a requested exposure time, and then performing a final $N_2$ reads, our new mode performs the reset and then executes the requested number of reads with a fixed exposure time delay between them, i.e., up-the-ramp (UTR) sampling. For our experiments the exposure time delay is set to 0 (i.e., we continually read the pixels). 

To support our hybrid image and guide window readout mode, we modified the full-frame clocking scheme to append window pixels to the end of each row.  We based our new clocking scheme on the Guide Window mode described in the HAWAII-1RG technical documentation as shown in Figure~\ref{fig:waveform}. We added an extra frame sync (FSYNCB) at the start of each windowed readout, as recommended by Roger Smith (private communication).

Our test code has a number of parameters that can be configured at runtime, including: (i) the total number of full-frame image reads, $N_\mathrm{reads}$; (ii) the window dimensions, $(N_\mathrm{wx}, N_\mathrm{wy})$; (iii) the corner coordinates of the windows, $(x_i,y_i)$, where $i=1..N_w$, and $N_w$ is the total number of windows; (iv) the full-frame reset interval, $I_\mathrm{reset}$ (i.e., the number of full-frame reads between full-frame resets); (v) the window reset interval, $I_\mathrm{winReset}$ (one in $I_\mathrm{winReset}$ window visits will both reset and read the pixels); and (vi) the window read interval, $I_\mathrm{winRead}$ (the window is only visited once every $I_\mathrm{winRead}$ full-frame rows, i.e., as a means to reduce the visit cadence).

The separate window read and reset intervals provide a great deal of flexibility to support the desired use cases. For example, bright guide stars may require more frequent resets than the full science image to avoid saturation (and as mentioned above, as a potential mechanism to mitigate localized persistence). It may be desirable to reduce the window read cadence to minimize the impact of self-heating.

When the window is not being read out, the window location is ``parked'' at the top-right of the detector (coordinates 2047, 2047), which causes subsequent pixels to run off the end of the detector and zero-values being returned instead of valid pixel data.  The zeros serve as placeholder pixels which still allows the API to properly format the image data according the image dimensions.

We send these parameters to the microcontroller using the ARC API.  We extended it to support new command packets with our additional parameters prior to commencing each exposure.  Our test application also allows us to sequence multiple exposures back-to-back, facilitating our experiments which probe certain parameters (e.g., window visit cadence). Setting the detector's window coordinates and enabling/disabling window mode requires serial commands to modify the HAWAII-2RG's internal registers. While a basic API call is provided by ARC to send such commands, we updated it to explicitly track their values and to simplify how they are updated following the documented register naming conventions.

We have been operating the HAWAII-2RG in single channel mode, with a 100\,kHz pixel clock. We typically allocate an additional 64 pixels to each row for window visits. Accounting for the window and additional overheads, the row period is 21.1\,ms, and a full-frame read takes 43.3\,s.

Each exposure is currently initiated using a blocking call from the program running on the host computer. This process results in a 16-bit raw data cube. The first two dimensions are typically 2112 columns (4$\times$2 reference pixels at the row ends, 2040 light-sensitive pixels along each row, and 64 arbitrarily-located window pixels), and 2048 rows (4$\times$2 reference pixels at the column ends, and 2040 light-sensitive pixels along each column). The third dimension enumerates read number. The cube is then written to a FITS file for future analysis.

A limitation of this system is that all of the reads must be stored in memory until the end of the sequence. The longest exposures that we were able to execute were about 100 full-frame reads, taking approximately 1\,hr 15\,min. A future goal for this system is to develop an asynchronous data acquisition mode. In this way data can be handled as soon as they arrive (e.g., via a callback function each time a packet of row data, including a full set of window pixels). This will allow us to both execute arbitrarily-long exposure sequences, and to process ODGW pixels in real-time.

We also encountered a number of issues for which we implemented workarounds, though in a production system we would want to more fully investigate their causes. Early in testing we noticed a strong vertical gradient in the initial read of the detector. Our conclusion was that the controller requires a substantial settling time when it is first powered-up, or when the bias is changed. We observed a bias change with an oscilloscope from 0\,V to 3.3\,V (typical operating point for the clock signals) and it took $\sim$10\,s to stabilize. Adding a 10\,s delay after starting the controller (and prior to any exposures) removed the gradient from the data. We found that the initial read of the detector as part of an exposure sequence is scrambled. It behaves as though some number of the initial row and column clocks are being missed and/or there is a synchronization issue with the video board. Alternating columns also appear to be interleaved. For our experiments we simply add a ``sacrifical read'' at the start of each exposure, as subsequent reads do not exhibit this problem (though this adds an initial $\sim$40\,s exposure overhead). When setting the video board offsets, we initially tried to center the full-well for typical pixels within the 16-bit range. However, we soon realized that something in the signal chain prior to the ADC was clipping values above $\sim$42000 ADUs. This corresponds to a reduction of $\sim$0.24\,V in usable dynamic range output by the detector (of a potential maximum of 0.68\,V), although it was still sufficient for our detector. Finally, we encountered sporadic stability issues. Occasionally it appears that some of the serial commands are missed when executed at the start of an exposure. The evidence is that the detector is misconfigured, resulting in runtime errors or otherwise bad pixel data.  The solution was to add a delay to the API code of an arbitrary length (we chose 3\,s) to give the controller time to finish each configuration command.  Despite this, we still periodically observe problems, such as the detector remaining in the default 4-channel output mode. Further delays do not seem to help, and we resort to re-executing the exposure in these instances. We suspect that we need to experiment further with the serial command waveforms; this problem may be related to the ``sacrifical frame'' issue mentioned above.

%
%

\subsection{Dark tests}
\label{sec:dark}

We first explored the impact of one or more windows on the detector without illumination. In general we precede each experiment with a ``dark'' exposure, during which the window mode is deactivated, but with extra time padding in the clock waveforms at full-frame row ends to match what is normally allotted to reading (typically 64) window pixels. In this way the overall integration time between full-frame reads is constant between the dark and subsequent exposures with the window running, and thus provides a good reference for the effects of the window on the full-frame reads.

\begin{figure}[hbt]
    \begin{center}
        \includegraphics[width=0.8\linewidth]{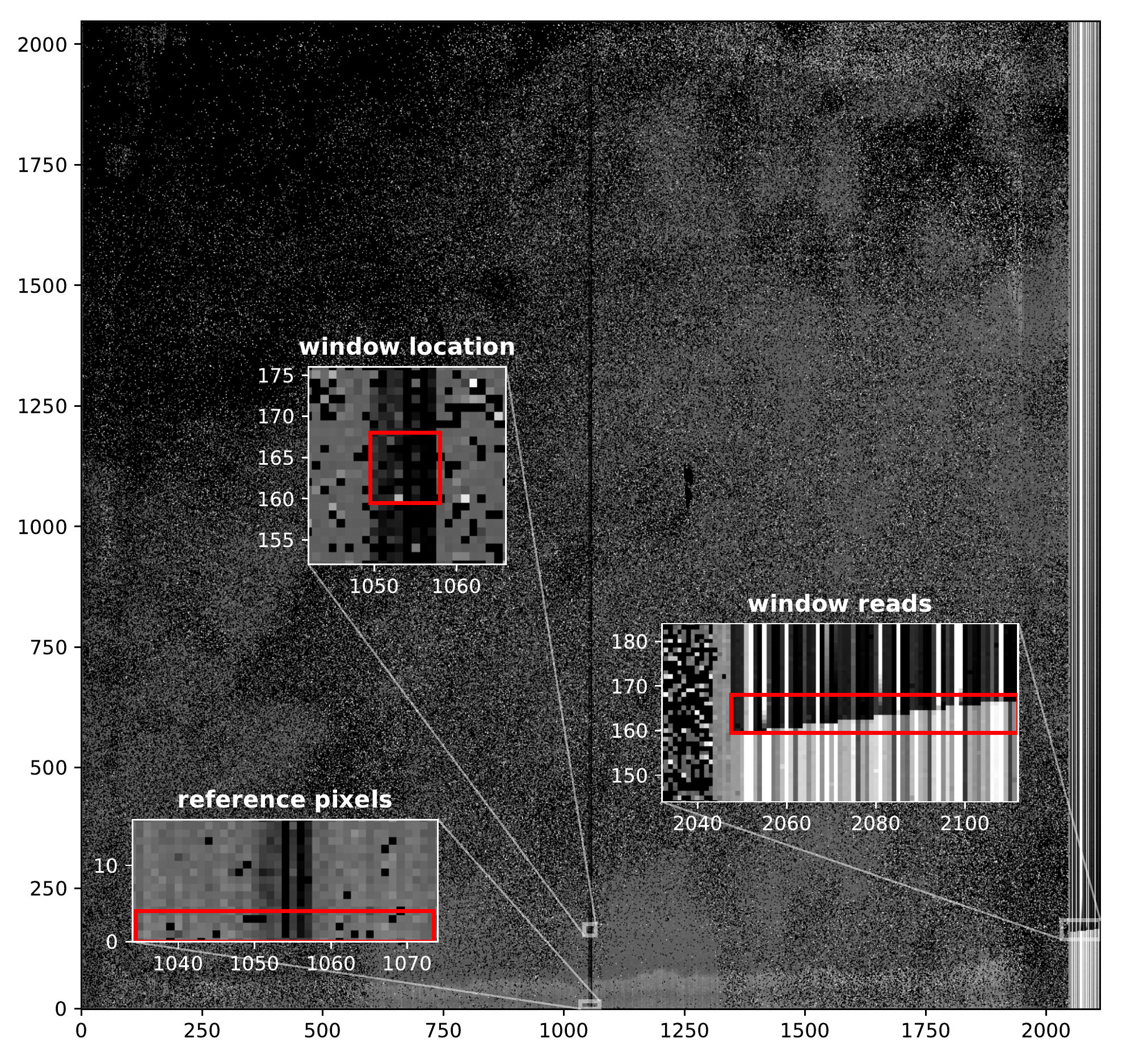}
    \end{center}  
    \caption{\label{fig:example}
    Illustration of the ODGW observing mode for our HAWAII-2RG chip. The displayed map shows the first read of a full-frame exposure with the ODGW reading continuously (each full set of window pixel reads are appended to the end of each row). The data have been dark subtracted (using an equivalent prior exposure with no window running as a reference). The window pixel reads in the rightmost 64 columns have their initial (reset) values, as measured from the first full-frame read, subtracted. Their values gradually increase up columns as they accumulate charge between resets. Since the initial window reads (below row 160) occur before those pixels have been reset, the effect of the rolling full-frame reset can be seen as progressively larger numbers of window pixels dropping to low values, until the entire window has been reset at row 167 in the right inset. This inset also shows four columns of reference pixels that lie between the full-frame image and the window reads. Finally, there is a prominent offset in the reset levels of all pixels in the full frame lying along columns coincident with the window. This effect extends to the reference pixels as shown in the lower inset. Note that approximately 50\% of the pixels on this chip are bad (particularly in the top-left quadrant), and are rendered in black. 
    }
\end{figure}

\begin{figure}[hbt]
    \begin{center}
        \includegraphics[width=0.7\linewidth]{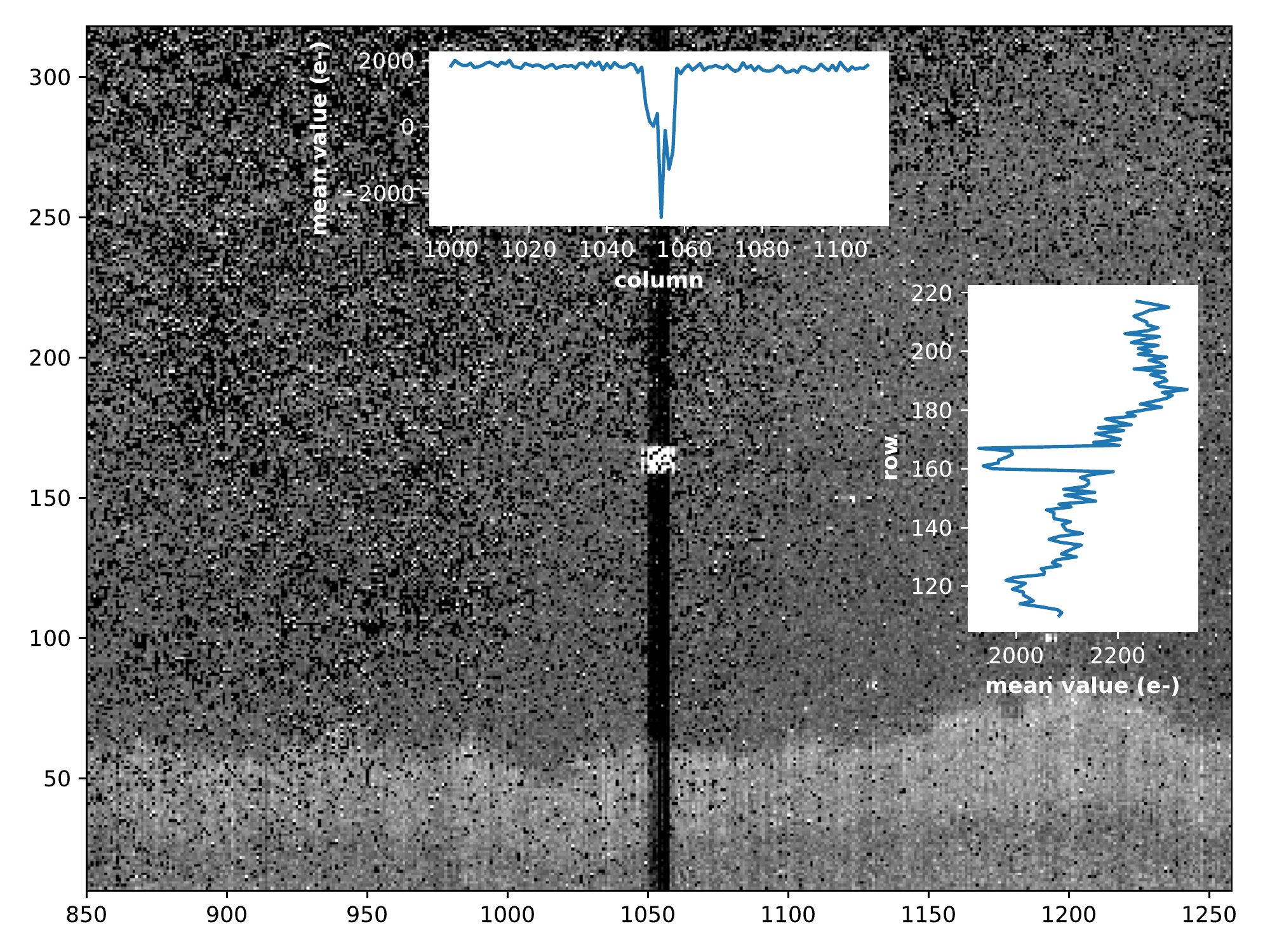}
    \end{center}  
    \caption{\label{fig:depression}
    Zoom-in of the region around the window for the same dark-subtracted data as Figure~\ref{fig:example}, but for the 20th read. Self-heating at the window location is now evident due to the more frequent visits. The top inset shows the magnitude of the offset along columns coincident with the window, of order $-3000$\,e- with respect to adjacent pixels. It is calculated by averaging rows of pixels away from the location of the window. Similarly the inset on the right shows a much fainter offset in rows coincident with the window of order $-150$\,e-, this time created by averaging columns away from the window.
    }
\end{figure}

\begin{figure}[hbt]
    \begin{center}
        \includegraphics[width=\linewidth]{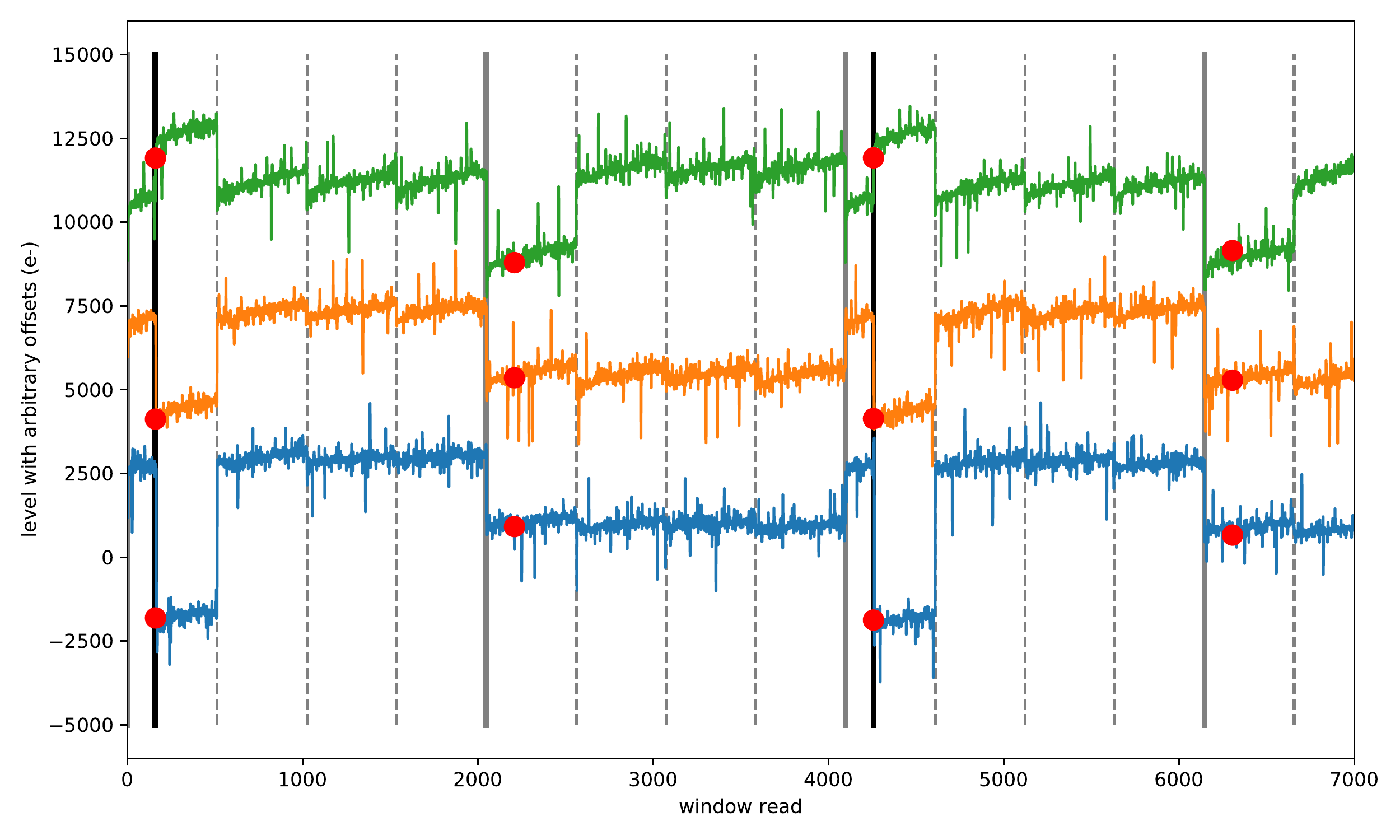}
    \end{center}  
    \caption{\label{fig:resets}
    Multiple window (colored lines) and full-frame (solid red circles) reads of several pixels within the window (every 2048 window reads, offset 160 from the start of each full-frame corresponding to the $y$-coordinate of the window), with a full-frame reset occurring once every two full-frame reads (black vertical lines, when the full-frame sweep reaches the window, every 4096 window reads), and window resets occurring every 512 window reads (dashed grey vertical lines). The different pixel traces have been separated vertically by arbitrary amounts. The levels resulting from the full-frame and window resets are clearly different, and repeatable. Additionally, window resets corresponding to the start of a new full-frame read (e.g., at 2048) results in a third, different level.
    }
\end{figure}

Figure~\ref{fig:example} illustrates a number of features of the ODGW mode using only minimal data reduction. In this example there is a single 8$\times$8 window being read at the end of each full-frame row near the bottom of the detector. The full detector is reset once as part of the first full-frame read, and then there are a total of 20 full-frame reads. Visits to the window pixels at the end of each row only read the values; there are no additional resets. Raw values written by the controller are inverted, so we first replace them with their 16-bit complement (i.e., $\mathrm{data} = 65535 - \mathrm{data}$). A prior dark exposure with the same number of reads is subtracted from the full-frame 2048$\times$2048 data cube. The final 64 columns of window pixel reads simply have the initial pixel values from the full-frame read subtracted (to roughly remove their reset levels). The only other processing step is to identify bad pixels ($\sim$50\%)-- this uses a masking procedure described later in this section. 

The first read of this dark-subtracted exposure clearly shows two main features: (i) a negative offset in the image along columns aligned with the window; and (ii) the rolling full-frame reset is seen passing through the window reads. Note that the offset is also visible in reference pixels at the top and bottom of detector. Offsets along columns have been observed in the past for HAWAII detectors operating in window mode \cite{smith2012}.

The 20th (final) read of this dark-subtracted exposure is shown in Figure~\ref{fig:depression}. The most prominent difference between this and the previous figure is that the pixels within the window have accumulated significantly more charge than their neighbors. This window self-heating is believed to be caused by the read-out integrated circuit (ROIC), and has been documented elsewhere\cite{smith2012,regan2020}. In addition, an inset at the top quantifies the depth of the negative offset along window columns, roughly 4000\,e-, established from the average of pixel rows away from the window. Finally, a similar exercise is used to illustrate the presence of a much fainter negative offset in rows coincident with the window, roughly 200\,e-. Negative row offsets have also been observed in the past due to the operation of windows \cite{bezawada2006,smith2012}.

To explore the impact of window resets on the full-frame pixels, and \emph{vice versa}, we conducted an experiment in which: (i) the full-frame reads include resets for every other read (i.e., read 0, 2, ...), and (ii) the window is read continuously, but is also reset every 512 full-frame rows. Time traces for several representative pixels from the 8$\times$8 window are shown in Figure~\ref{fig:resets}. The most striking feature of this plot is that the reset level for a given pixel differs depending on whether it occurs as part of the full-frame line reset, or within the window. The sign of this difference also appears to be a property of the pixel (e.g., the full-frame row reset is deeper for the blue and orange pixels, but shallower for the green pixel). The other curious feature is that the first reset after a new full-frame read begins, \emph{for which there is no accompanying full-frame row reset} (e.g., at window read 2048 in the plot), results in a third, different reset level. While we do not propose a physical model for these three reset states, we note that the pattern is repeatable and should be removable during data reduction.

\subsubsection{Data reduction}

A more sophisticated data reduction procedure was developed for the remaining experiments in this paper. The following points summarize the main steps, and resulting data products:

\begin{enumerate}
    \item Calculate 16-bit complement of data (i.e., data = 65535 - data) so that increasing values correspond to greater charge.
    \item Separate full-frame 2048$\times$2048 reads and window (e.g., 8$\times$8) reads into different data cubes.
    \item Fit lines (slope=flux, and offset=reset level) to the sequence of reads between resets for each pixel, in both the full-frame and window data cubes.
    \item Establish a good pixel mask by rejecting outliers via iterative sigma-clipping in the full-frame flux (3-$\sigma$) and reset level (2.5-$\sigma$) maps. The reference pixel masking is performed independently of the light-sensitive pixels (and only small numbers are rejected using this technique).
    \item Calculate residual data cubes (subtracting the linear flux model from each pixel ramp).
\end{enumerate}

\begin{figure}
    \begin{center}
        \includegraphics[width=\linewidth]{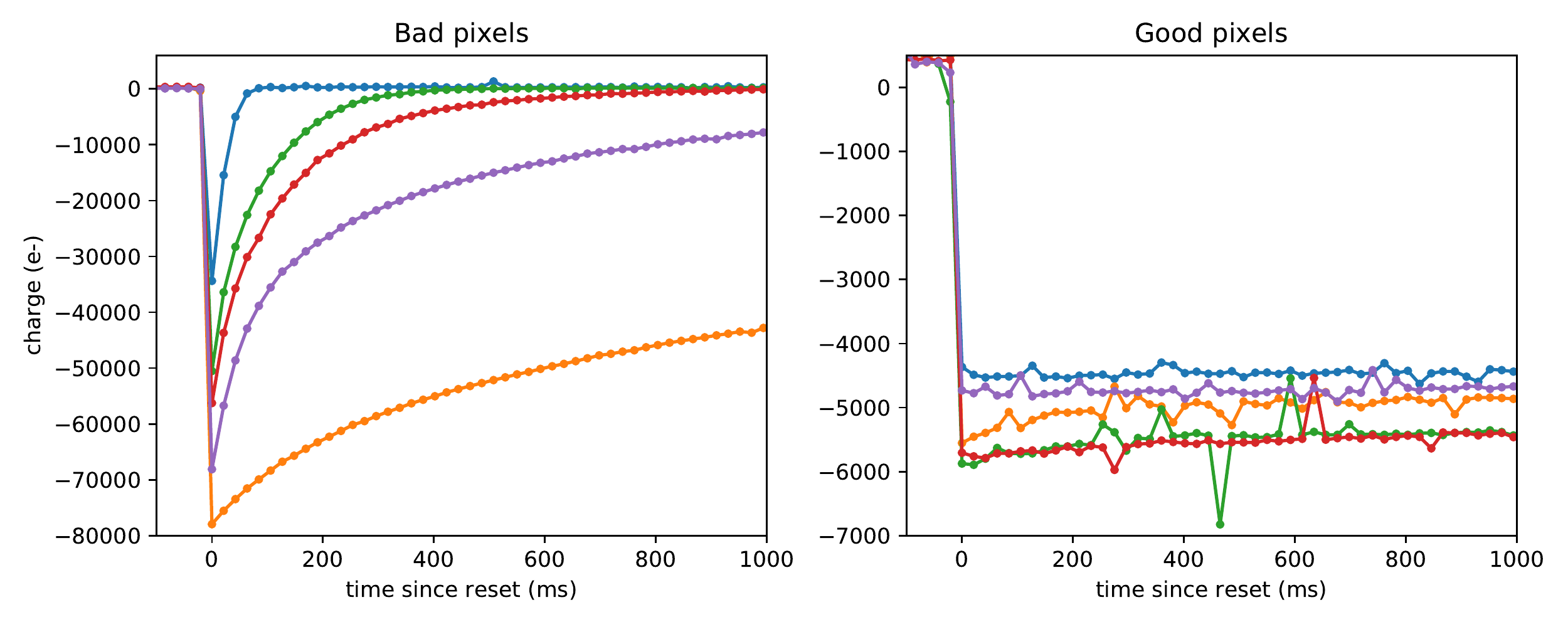}
    \end{center}  
    \caption{\label{fig:pixramps}
    Comparison of bad pixel time traces (left) ,as flagged by our masking algorithm, with good pixel traces (right) after a reset, as observed in the window pixel reads.
    }
\end{figure}

The pixel mask is typically calculated from a dark exposure so that it will not be influenced by structures induced by the window, and can then be applied to subsequent exposures. Approximately 50\% of the pixels are flagged as ``bad'' using this procedure. Figure~\ref{fig:pixramps} illustrates the difference between good and bad pixels, using the faster reads of pixels within the window to more finely sample their behavior over time. Bad pixels typically saturate within 10s of milliseconds after a reset, whereas good pixels exhibit the expected linear growth after a reset.

The right-hand plot in this figure also shows the settling behavior (typically referred to as the ``reset anomaly'') in good pixels immediately after a reset (e.g., the brief overshoot in the orange pixel trace which becomes linear about ~150--200\,ms later). Since the first full-frame pixel reads occur shortly after the row reset, their values can be significantly impacted by the reset anomaly. We therefore ignore the first read in the line-fitting procedure when calculating full-frame flux and offset maps (and we can similarly ignore the first few samples if performing linear fits to faster window pixel reads).

\subsubsection{Varying window visit cadence and number of windows}

\begin{figure}[hbt]
    \begin{center}
        \includegraphics[width=\linewidth]{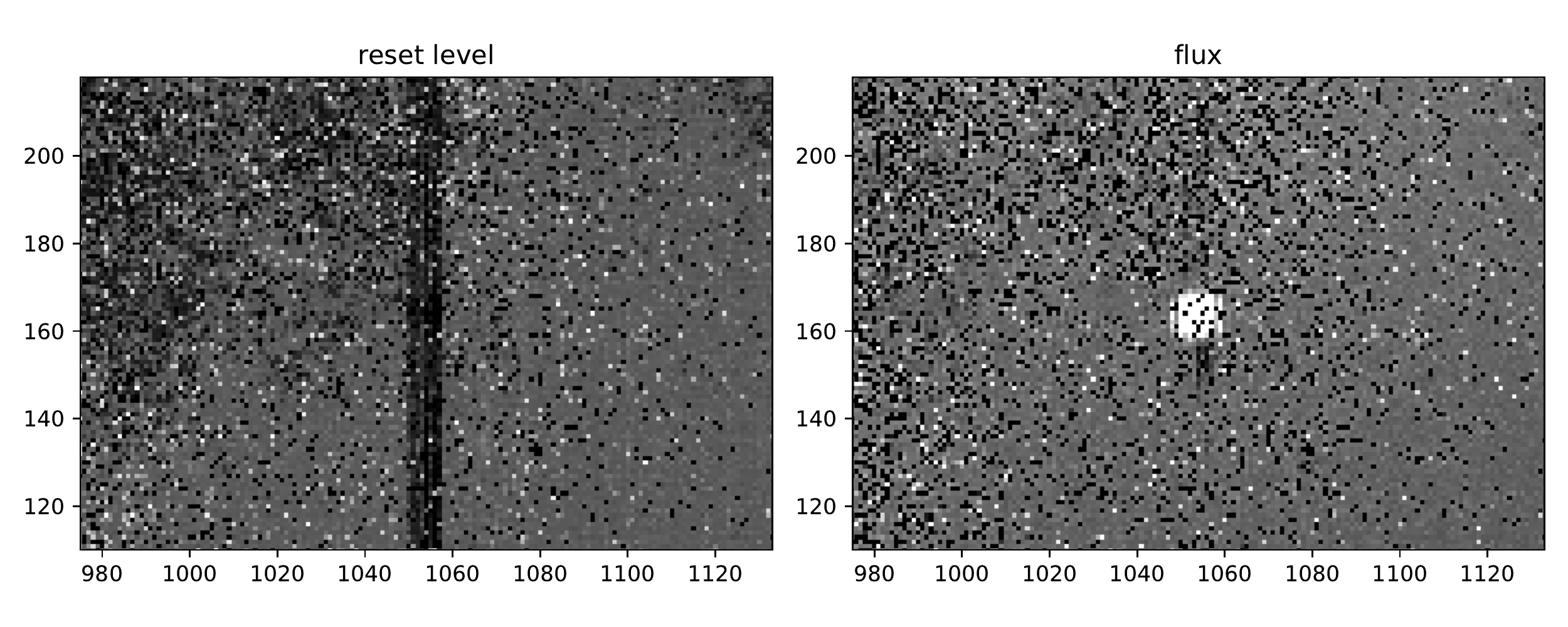}
    \end{center}  
    \caption{\label{fig:read0}
    Reset level and flux maps inferred from data in which the window was being read (but not reset) every row, by fitting a line to the individual pixel ramps.
    }
\end{figure}

We conducted several tests to explore how different window operational scenarios might impact the main science image. We varied the window visit cadence (with and without the window reset running), and we also operated multiple windows at different locations on the chip. In all cases we precede the experiments with a dark exposure for reference (without the window(s) running), and subtracted the dark from the results.

Figure~\ref{fig:read0} shows the reset level and flux maps that result from the line fitting procedure when reading the 8$\times$8 window continuously (no window resets). It can be directly compared to figure Figure~\ref{fig:depression}, and shows how the model fitting cleanly separates the column offset into a component of the reset level map (i.e., it appears as a bias), while self-heating appears in the flux map (i.e., it increases the slope of the line).


We next experimented with varying the cadence of the window reads, from once at the end of each full-frame row as in the previous examples, to once every 4, 8, and 16 rows. The corresponding window visit rates were 47.3\,Hz, 11.8\,Hz, 5.91\,Hz, and 2.95\,Hz. We found that the mean excess flux in the window pixels varied $\sim$linearly from 11\,e-/s at the highest rate, to 0.6\,e-/s at the lowest rate. The neighboring pixels beyond the edge of the window pixels showed a mean excess flux of $\sim$1\,e-/s at the highest rate. The data were too noisy to break down the excess flux by location in and around the window, particularly at the lower visit rates.

\begin{figure}
    \begin{center}
        \includegraphics[width=\linewidth]{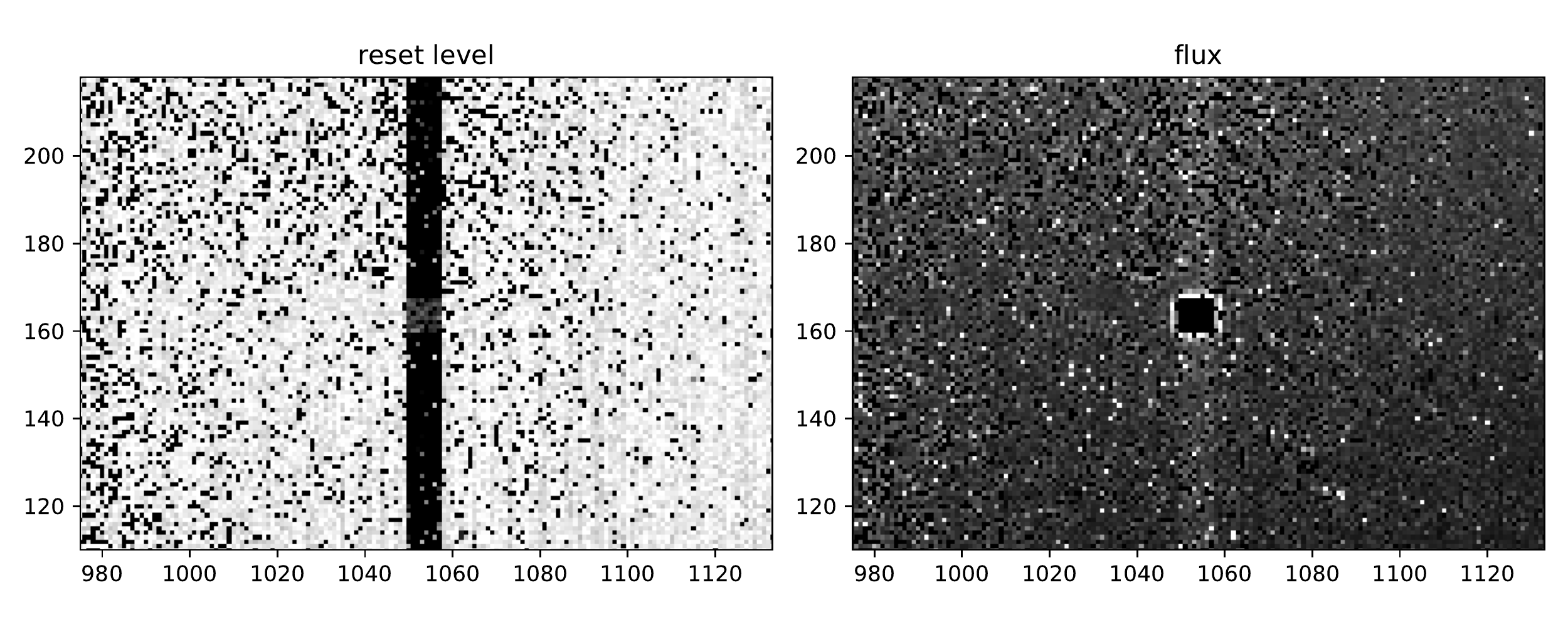}
    \end{center}  
    \caption{\label{fig:rst0}
    Inferred reset level and flux maps when the window is being reset every row. The reset level of pixels within the window are lower when compared to adjacent columns, but higher than other pixels outside of the window but along the same columns. The flux map shows a glow extending to pixels just outside the window, and even fainter excess flux along columns.
    }
\end{figure}

\begin{figure}
    \begin{center}
        \includegraphics[width=\linewidth]{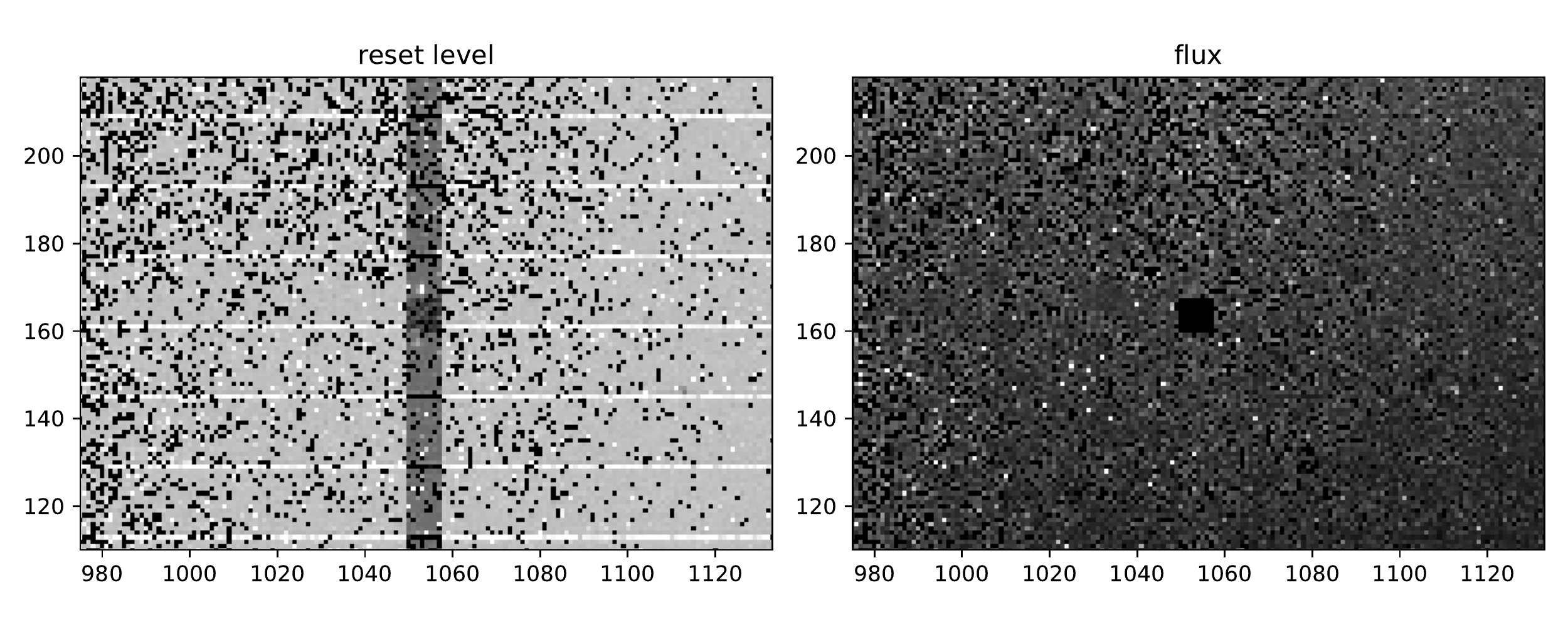}
    \end{center}  
    \caption{\label{fig:rst16}
    Same as Figure~\ref{fig:rst0}, but now only reading  and resetting the window once every 16 full-frame rows. The bright horizontal stripes in the reset level map occur on rows immediately following window resets.
    }
\end{figure}

We then repeated these tests while resetting the window each time it was visited. Figure~\ref{fig:rst0} shows the results when resetting every row (47.3\,Hz). Compared to the previous plot, the reset level differences along columns coincident with the window are more pronounced ($\sim-9500$\,eV compared to adjacent pixels). Interestingly, the negative offset of the window pixels is actually shallower ($\sim-6000$\,eV). The flux map shows evidence both for blooming extending past the window pixels (the immediately adjacent pixels exhibit a mean excess flux of $\sim$8\,e-/s), but also a faint excess flux along columns (of order $\sim0.2$\,e-/s).

In Figure~\ref{fig:rst16} the window is now only reset every 16 rows (2.95\,Hz). While self-heating artifacts are substantially reduced in the flux map, there is a strong pattern in the reset level map correlated with the rows where the window resets occur. In general, the row of pixels immediately following a row where the window was reset exhibits an elevated signal. We also note three different reset levels along window columns: (a) rows outside of the window that follow window resets are the deepest; (b) pixels within the window are slightly shallower; and (c) all remaining pixels in these columns have the shallowest resets (though they are still deeper than the majority of pixels in the rest of the detector).

\begin{figure}[hbt]
    \begin{center}
        \includegraphics[width=\linewidth]{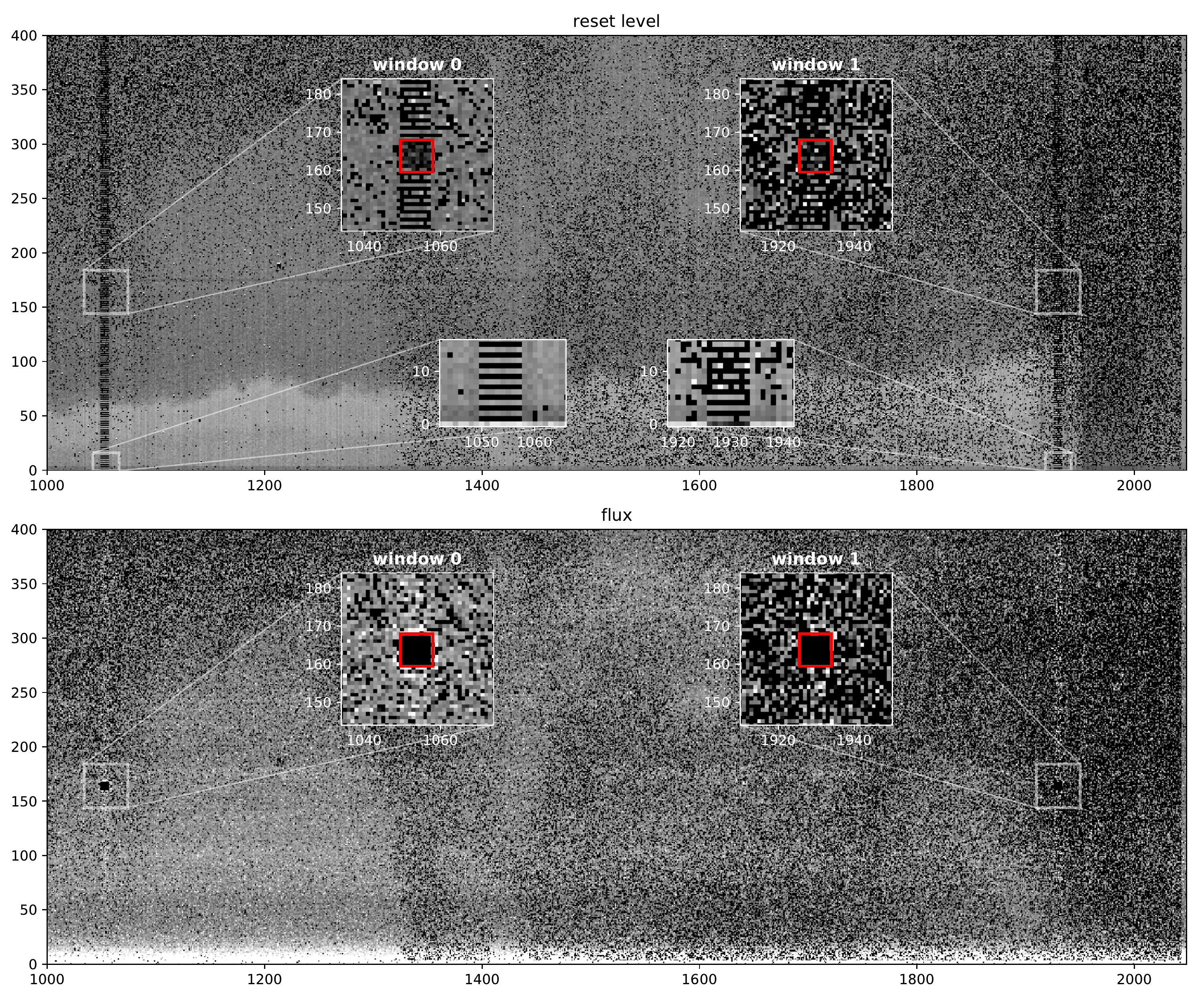}
    \end{center}  
    \caption{\label{fig:multiSameRows}
    An experiment in which we ran two windows on the same rows, both being held in continuous reset. The resuls are dark-subtracted. The top figure shows the reset map, and the bottom figure the flux map. Each window is addressed at the end of alternative full-frame row readouts. The lower insets in the top map shows how the negative reset offsets above and below the windows occur on alternating rows. These patterns a largely removed from the linear fits to the ramps as shown in the flux map.
    }
\end{figure}

Finally, we ran several experiments where we placed two or more windows at different locations. Figure~\ref{fig:multiSameRows} shows one example where two windows were placed along the same rows, but separated by a large number of columns. Only a single window can be visited at a time, so the per-window rate is now reduced to 23.6\,Hz, with window 0 being reset on even rows, and window 1 on odd rows. Comparing with Figure~\ref{fig:rst16}, the horizontal stripes in the reset map are now restricted to the columns coincident with the windows. We also note that the stripes occur on alternating rows for the two windows.

Again, we conclude that these patterns observed in the reset level, though complex, should be possible to calibrate out as they appear to be stable on the timescale of our test exposures (typically 15\,min).

%
%

\subsection{Persistence tests}
\label{sec:led}

\begin{figure}
    \begin{center}
        \includegraphics[width=\linewidth]{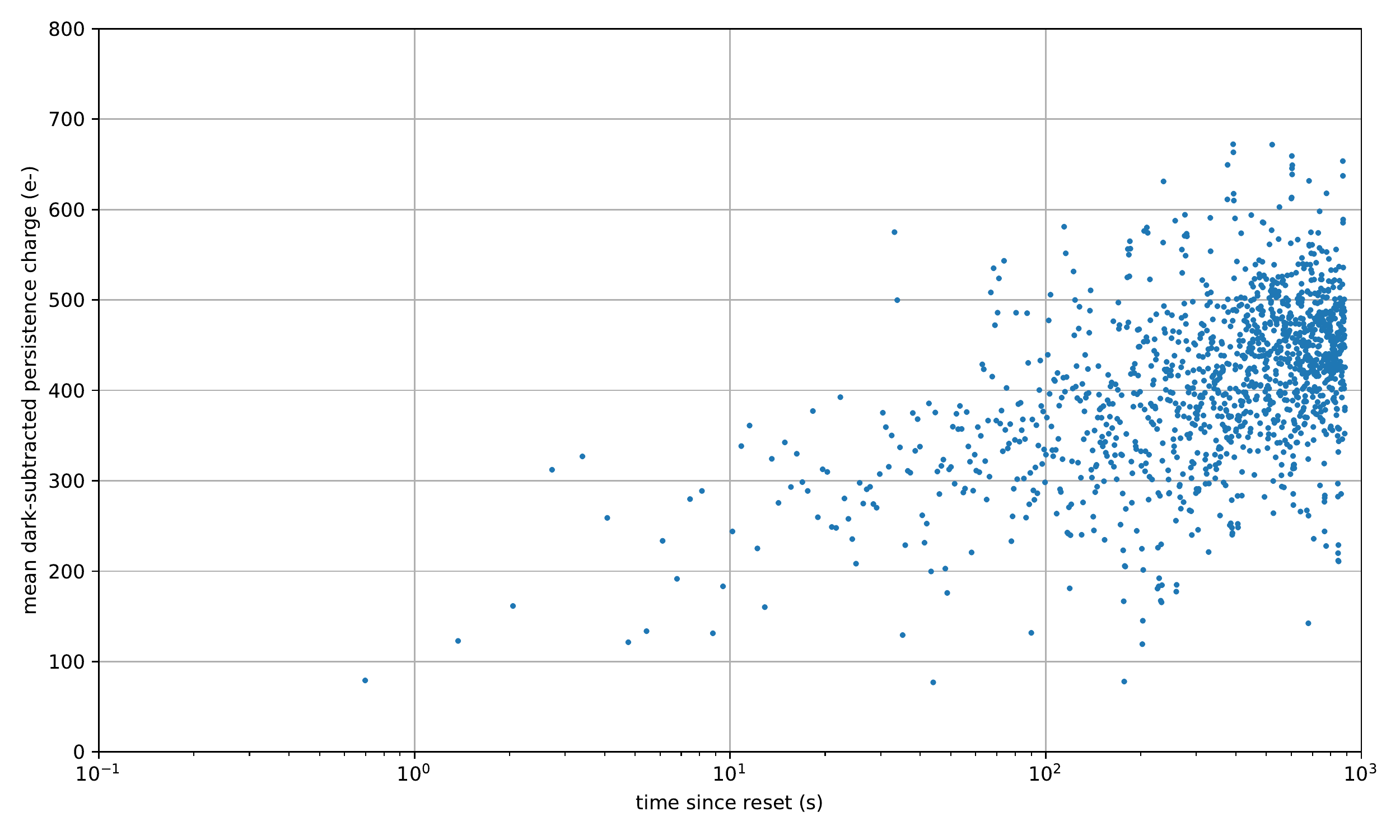}
    \end{center}  
    \caption{\label{fig:soakPix}
    Mean dark-subtracted 8$\times$8 window pixel reads after a reset following a 2-s flash and 120-s soak with an LED. The window is visited at 1.47\,Hz.
    }
\end{figure}

Next we experimented with illuminating the detector using the LED. Our initial goal was to verify that we could induce and measure persistence in our detector similar to the HAWAII-2RG experiments of Tulloch 2018\cite{tulloch2018}. The basic idea, following the trap model for persistence\cite{smith2008}, is that the persistent dark current following illumination is caused by the slow release of charges that have been captured by traps. The number of trapped charges is thought to be a complex function of the incident flux, accumulated charge within the pixels, and the time between resets.

To facilitate these tests, the LED was mounted in the cryostat with a simple pinhole replacing the lens assembly to create a broad illumination pattern. It was connected to one of the timing signal outputs of the controller and the code was updated to enable a special illumination exposure. After an initial reset and read of the detector, the LED is activated for a configurable ``flash'' time, $T_\mathrm{flash}$. Next, all of the timing signals are halted for a ``soak'' time, $T_\mathrm{soak}$ (to allow more charges to become trapped). The sequence concludes with a read to measure how much charge was accumulated during the flash. This special exposure is immediately followed by a normal exposure in which the detector is first reset, and then read continuously (both the full-frame, and window pixels) in order to track any time-varying dark current.

Prior to our experiments at the start of the day (before exposing the detector to any light) we performed a dark exposure. We then performed the flash, soak, reset and read sequence as described above. We typically flashed for 2\,s which would saturate the central region of the detector, falling to roughly 1/3 of the full-well depth by the edges. We placed a single window over a patch of good pixels near the centre of the detector and read it once every 32 rows (1.47\,Hz), to balance speed of the readout with self-heating that might affect any observed excess dark current. Figure~\ref{fig:soakPix} shows an estimate of the mean excess charge in window pixels after soaking for 120\,s. The first read of each pixel ramp following the reset is subtracted to approximately remove the reset level. The equivalent process is applied to the dark exposure, and the difference is taken. Finally, the average value of the pixel reads is plotted. While the result is extremely noisy, the general trend appears to be consistent with Tulloch (2018)\cite{tulloch2018} for a comparable soak time. However, when we repeated this test for longer soak times of 240\,s, 480\,s and 2000\,s, we did not see a convincing trend of increasing dark current (in fact, it appeared to drop at 240\,s, and then returned to the same level for a 480\,s soak).

We decided that the system would need to be refined before we can attempt to measure the potentially subtle effect of a continuously resetting window reliably. The individual pixel reads are extremely noisy, and it is presently unclear whether it has to do with the general poor quality of the detector, noise in the data acquisition system, or whether we simply require a more sophisticated data reduction algorithm. It is also clear from the plot above that finer sampling is needed. In retrospect we should simply have run the window at the maximum rate of 47.3\,Hz (and ensured that the dark exposure was executed in the same way so that any self-heating would be accounted for). For the next run of this experiment we will also ensure that the temperature control system is operating.

\section{Summary and future work}

We successfully implemented an On-Detector Guide Window mode for a HAWAII-2RG detector using the new Astronomical Research Cameras (ARC) Gen-4 controller, demonstrating the ability to conduct fast reads (and resets) of the window in parallel with the slower full-frame science reads. This capability is a key enabling technology that will be needed to meet the ambitious adaptive optics performance goals set for new science instruments on existing 8--10\,m class telescopes, and future extremely large telescopes. With a single $8\times8$ window, a single-channel output, and running in the slow (100\,kHz pixel clock) mode for the detector, we achieved peak window visit rates of 47.3\,Hz (0.021\,s/read) while full-frame reads occur at 0.023\,Hz (43.3\,s/read). It is possible to independently configure the window visit and window reset rates at multiples of the window visit period. Our system can also visit multiple windows at different coordinates, though the peak visit rate is reduced by the number of windows as only one may be visited at a time. The window visit rates could be increased by using multiple video channels and/or running in one of the faster output modes. We conducted an array of experiments, including varying the window read and reset cadences, and placing them at different locations on the detector. Transient self-heating artifacts (increased flux in and near the location of the window), and complex offset patterns were observed. The latter appear to be caused by different levels in the full-frame line resets as compared with resets implemented during the window visits. Fortunately these reset patterns are stable and repeatable. A simple linear fit to the individual pixel ramps was demonstrated to effectively separate the reset (offset) effects from the measured fluxes (slope) in the science image. Finally, we attempted to induce persistence in our detector by flashing it with an LED, with the intention of holding a window within the illuminated region in reset to minimize long-term persistence. While the preliminary results indicate that there is persistence at a plausible level, the data are extremely noisy and we did not attempt to measure the impact of frequent window resets at this time.

In the second phase of this project in the autumn, there are several improvements that we plan to make. First, our system currently collects all of the data and writes it to disk once an exposing sequence is complete. We plan to work with ARC to develop an asynchronous acquisition mode that will allow us to intercept rows of data as they are sent to the host computer from the controller so that the window data can be used in real time. We also hope to replace our highly-degraded HAWAII-2RG with a JWST flight spare that should not have suffered the indium diffusion problem. This should allow us to make improved measurements of the effects described in this paper. We also plan to revisit the persistence experiments. Finally, we have been collaborating with S. Sivanandam at the University of Toronto (GIRMOS PI) who has been investigating ODGWs using an alternative controller and a HAWAII-4RG that will be used as the GIRMOS imager. We hope to re-run our tests using this same detector so that we can directly compare the results.

\acknowledgments

The authors wish to thank Suresh Sivanandam (UofT) for valuable ongoing discussions while we were developing and analyzing results from our test system. Bob Leach and Scott Streit from Astronomical Research Cameras provided frequent assistance with the use of their new controller. We also thank Tim Greffe and Roger Smith (Caltech) for providing us with results from their own ODGW experiment, and suggestions for how to configure our detector. Finally, we thank Rick Murowinski and Alan McConnachie (HAA) for discussions that helped guide this work. This project was supported by a New Beginnings grant from National Research Council Canada.

\bibliography{refs} 

\begin{thebibliography}{10}

\bibitem{larkin2020}
{Larkin}, J.~E., {Wright}, S.~A., {Chisholm}, E.~M., {Andersen}, D., {Dekany},
  R.~G., {Dunn}, J.~S., {Hayano}, Y., {Kupke}, R., {Smith}, R., {Suzuki}, R.,
  {Weber}, R.~W., and {Zhang}, K., ``{The Infrared Imaging Spectrograph (IRIS)
  for TMT: instrument overview},'' in [{\em Society of Photo-Optical
  Instrumentation Engineers (SPIE) Conference
  Series}{\nolinebreak\hspace{0.1em}]},  {\em Society of Photo-Optical
  Instrumentation Engineers (SPIE) Conference Series} {\bf 11447},  114471Y
  (Dec. 2020).

\bibitem{dunn2016}
{Dunn}, J., {Andersen}, D., {Chapin}, E., {Reshetov}, V., {Wierzbicki}, R.,
  {Herriot}, G., {Chalmer}, D., {Isbrucker}, V., {Larkin}, J.~E., {Moore},
  A.~M., and {Suzuki}, R., ``{The Infrared Imaging Spectrograph (IRIS) for TMT:
  multi-tiered wavefront measurements and novel mechanical design},'' in [{\em
  Ground-based and Airborne Instrumentation for Astronomy
  VI}{\nolinebreak\hspace{0.1em}]},  {Evans}, C.~J., {Simard}, L., and
  {Takami}, H., eds., {\em Society of Photo-Optical Instrumentation Engineers
  (SPIE) Conference Series} {\bf 9908},  9908A9 (Aug. 2016).

\bibitem{chapman2018}
{Chapman}, S.~C., {Sivanadam}, S., {Andersen}, D., {Bradley}, C., {Correia},
  C., {Lamb}, M., {Lardiere}, O., {Ross}, C., {Sivo}, G., and {Veran}, J.-P.,
  ``{The multi-object adaptive optics system for the GIRMOS spectrograph on
  Gemini-South},'' in [{\em Adaptive Optics Systems
  VI}{\nolinebreak\hspace{0.1em}]},  {Close}, L.~M., {Schreiber}, L., and
  {Schmidt}, D., eds., {\em Society of Photo-Optical Instrumentation Engineers
  (SPIE) Conference Series} {\bf 10703},  107031K (July 2018).

\bibitem{loic2005}
Albert, L., Riopel, M., Teeple, D., Ward, J., and Barrick, G., ``{The on-chip
  guiding system of the wide-field infrared camera at CFHT},'' in [{\em
  Infrared and Photoelectronic Imagers and Detector
  Devices}{\nolinebreak\hspace{0.1em}]},  Longshore, R.~E., ed.,  {\bf 5881},
  71 -- 78, International Society for Optics and Photonics, SPIE (2005).

\bibitem{baril2006}
{Baril}, M.~R., {Ward}, J., {Teeple}, D., {Barrick}, G., {Albert}, L.,
  {Riopel}, M., and {Wang}, S.-Y., ``{CFHT-WIRCam: interlaced science and
  guiding readout with the Hawaii-2RG IR sensor},'' in [{\em Society of
  Photo-Optical Instrumentation Engineers (SPIE) Conference
  Series}{\nolinebreak\hspace{0.1em}]},  {McLean}, I.~S. and {Iye}, M., eds.,
  {\em Society of Photo-Optical Instrumentation Engineers (SPIE) Conference
  Series} {\bf 6269},  62690Z (June 2006).

\bibitem{vermeulen2006}
{Vermeulen}, T., {Teeple}, D., {Mahoney}, B., {Thomas}, J., {Albert}, L.,
  {Martin}, P., {Forveille}, T., and {Yan}, C.-H., ``{CFHT WIRCam software
  architecture and implementation},'' in [{\em Society of Photo-Optical
  Instrumentation Engineers (SPIE) Conference
  Series}{\nolinebreak\hspace{0.1em}]},  {Lewis}, H. and {Bridger}, A., eds.,
  {\em Society of Photo-Optical Instrumentation Engineers (SPIE) Conference
  Series} {\bf 6274},  62740J (June 2006).

\bibitem{boss2009}
{Boss}, A.~P., {Weinberger}, A.~J., {Anglada-Escud{\'e}}, G., {Thompson},
  I.~B., {Burley}, G., {Birk}, C., {Pravdo}, S.~H., {Shaklan}, S.~B.,
  {Gatewood}, G.~D., {Majewski}, S.~R., and {Patterson}, R.~J., ``{The Carnegie
  Astrometric Planet Search Program},'' {\em \pasp}~{\bf 121},  1218 (Nov.
  2009).

\bibitem{young2012}
{Young}, P.~J., {McGregor}, P., {van Harmelen}, J., and {Neichel}, B., ``{Using
  ODGWs with GSAOI: software and firmware implementation challenges},'' in
  [{\em Software and Cyberinfrastructure for Astronomy
  II}{\nolinebreak\hspace{0.1em}]},  {Radziwill}, N.~M. and {Chiozzi}, G.,
  eds., {\em Society of Photo-Optical Instrumentation Engineers (SPIE)
  Conference Series} {\bf 8451},  845124 (Sept. 2012).

\bibitem{bezawada2006}
{Bezawada}, N. and {Ives}, D., ``{High-speed multiple window readout of
  Hawaii-1RG detector for a radial velocity experiment},'' in [{\em Society of
  Photo-Optical Instrumentation Engineers (SPIE) Conference
  Series}{\nolinebreak\hspace{0.1em}]},  {Dorn}, D.~A. and {Holland}, A.~D.,
  eds., {\em Society of Photo-Optical Instrumentation Engineers (SPIE)
  Conference Series} {\bf 6276},  62760O (June 2006).

\bibitem{smith2012}
{Smith}, R.~M. and {Hale}, D., ``{Read noise for a 2.5{\ensuremath{\mu}}m
  cutoff Teledyne H2RG at 1-1000Hz frame rates},'' in [{\em High Energy,
  Optical, and Infrared Detectors for Astronomy
  V}{\nolinebreak\hspace{0.1em}]},  {Holland}, A.~D. and {Beletic}, J.~W.,
  eds., {\em Society of Photo-Optical Instrumentation Engineers (SPIE)
  Conference Series} {\bf 8453},  84530Y (July 2012).

\bibitem{atwood2022}
{Atwood}, J., {Crane}, J., {McConnachie}, A., {Reshetov}, V., {Lothrop}, J.,
  and {Lamb}, M., ``{A near-IR imager for the Gemini InfraRed Multi-Object
  Spectrograph (GIRMOS)},'' {\em \spie} (July 2022).

\bibitem{rauscher2012}
Rauscher, B.~J., Stahle, C., Hill, R.~J., Greenhou~se, M., Beletic, J., Babu,
  S., Blake, P., Cleve~land, K., Cofie, E., Eegholm, B., Engelbracht, C.~W.,
  Ha~ll, D. N.~B., Hoffman, A., Jeffers, B., Jhabvala, Christine an d~Kimble,
  R.~A., Kohn, S., Kopp, R., Lee, D., Leidecker, H., Lindler, D., McMurray,
  R.~E., Misselt, K., Mott, D.~B., Ohl, R., Pipher, J.~L., Piquette, E., Polis,
  D., Pontius, J., Rieke, M., Smith, R., Tennant, W.~E., Wang, L., Wen, Y.,
  Willmer, C. N.~A., and Zandian, M., ``Commentary: Jwst near-infrared detector
  degradation— finding the problem, fixing the problem, and moving forward,''
  {\em AIP Advances}~{\bf 2}(2),  021901 (2012).

\bibitem{regan2020}
{Regan}, M.~W. and {Bergeron}, L.~E., ``{Zero dark current in H2RG detectors:
  it is all multiplexer glow},'' {\em Journal of Astronomical Telescopes,
  Instruments, and Systems}~{\bf 6},  016001 (Jan. 2020).

\bibitem{tulloch2018}
{Tulloch}, S., ``{Persistence Characterisation of teledyne H2RG detectors},''
  {\em arXiv e-prints} ,  arXiv:1807.05217 (July 2018).

\bibitem{smith2008}
{Smith}, R.~M., {Zavodny}, M., {Rahmer}, G., and {Bonati}, M., ``{A theory for
  image persistence in HgCdTe photodiodes},'' in [{\em High Energy, Optical,
  and Infrared Detectors for Astronomy III}{\nolinebreak\hspace{0.1em}]},
  {Dorn}, D.~A. and {Holland}, A.~D., eds., {\em Society of Photo-Optical
  Instrumentation Engineers (SPIE) Conference Series} {\bf 7021},  70210J (July
  2008).

\end{thebibliography}
\bibliographystyle{spiebib} 

\end{document}